\documentclass[preprint]{aastex}
\usepackage{graphicx} 
\usepackage{figcaps}
\usepackage{balance}

\shorttitle{Long Nanoflare Storm}
\shortauthors{Mulu-Moore et. al}

\figcapsoff

\begin{document}


\title{Can a Long Nanoflare Storm Explain the Observed Emission Measure 
Distributions in Active Region Cores?}

\author{Fana M. Mulu-Moore\altaffilmark{1}, Amy R. Winebarger\altaffilmark{1}, 
and Harry P. Warren\altaffilmark{2}}
\altaffiltext{1}{NASA Marshall Space Flight Center, VP 62, Huntsville, AL 35812; 
fanamariam.mulumoore@nasa.gov}
\altaffiltext{2}{Space Science Division, Naval Research 
Laboratory, Washington, DC 20375}


\begin{abstract}
All theories that attempt to explain the heating of the high temperature plasma 
observed in the solar corona are based on short bursts of energy. The
intensities and velocities measured in the cores of quiescent active regions, 
however, can be steady over many hours of observation. One heating scenario that 
has been proposed to reconcile such observations with impulsive heating models 
is the ``long nanoflare storm,'' where short duration heating events occur 
infrequently on many sub-resolutions strands; the emission of the strands is 
then averaged together to explain the observed steady structures. In this 
Letter, we examine the emission measure distribution predicted for such a long 
nanoflare storm by modeling an arcade of strands in an active region core. 
Comparisons of the computed emission measure distributions with recent 
observations indicate that that the long nanoflare storm scenario implies 
greater than 5 times more 1 MK emission than is actually observed for all 
plausible combinations of loop lengths, heating rates, and abundances. We 
conjecture that if the plasma had ``super coronal'' abundances, the model may be 
able to match the observations at low temperatures.
\end{abstract}


\section{Introduction}
Determining the mechanism that heats the solar upper atmosphere is one of the 
central goals in observing the solar corona.  Coronal heating theories predict
short bursts of energy \citep{klimchuk2006}; these energy releases 
are often called ``nanoflares,'' though they are not necessarily caused by
small-scale magnetic reconnection events (i.e. Parker's nanoflares; \citealt{parker1988}). 
One interpretation of this type of heating
has been discussed in a series of papers by Cargill and Klimchuk
\citep{cargill1994,cargill1997,cargill2004,klimchuk2008,klimchuk2009}.  They describe 
nanoflare heating as a single short-lived heating event that occurs 
\emph{infrequently} on an elemental strand in the corona; the density and 
temperature of the plasma associated with the strand then evolve without any 
additional heating events.  (We adopt the definition of ``strand'' as a 
fundamental flux tube in the corona and ``loop'' as an observed coronal 
structure. A loop may be formed of a single strand, which would imply the corona 
is being resolved, or many, sub-resolution strands.)  The evolution and other 
properties of warm EUV loops have been shown to be consistent with the Cargill 
and Klimchuk nanoflare model 
\citep[e.g.,][]{warren2002b,winebarger2003a,warren2003}.  In these loops, the 
nanoflares occur within a narrow heating envelope (a so-called ``short nanoflare 
storm,'' \citealt{klimchuk2009}) on a few sub-resolution strands, then the 
strands, and apparent loop, evolve. 

In the core of active regions, however, the average apex intensity 
\citep{warren2010a}, footpoint intensity \citep{antiochos2003}, and Doppler and 
non-thermal velocities \citep{brooks2009} are often found to be steady (with 
variations of $\pm 20$\%) over several hours of observation, prompting some to 
suggest that the heating in active region cores is effectively steady, i.e., 
heating events occur {\it frequently} on a single strand in the corona 
\citep{schrijver2004,warren2006,warren2007,lundquist2008a,winebarger2008,winebarger2011}.  
\cite{klimchuk2009} has instead suggested that these loops are heated by 
``long nanoflare storms'', i.e. the loops are bundles of many strands and each 
strand is heated at a different time.  After a single heating event, the plasma 
along the strand is allowed to cool completely with no additional heating events 
on the strand.  At any given time in a loop's evolution, it is composed of 
strands in all the different stages of the heating and cooling cycle so the 
dynamic properties of the plasma are ``averaged out'' in the observed loop. For 
this heating scenario, the observed loops would have some of the same properties 
as a frequently heated loop, such as steady intensity and velocity.  

The purpose of this Letter is to test whether the emission measure distribution 
predicted by the long nanoflare storm model is consistent with the observed 
emission measure distributions in active region cores. We first calculate the 
expected temperature and density evolution for nine representative strands with 
a range of lengths and heating rates typical to active region cores using a 
one-dimensional hydrodynamic simulation. We complete the simulations twice, once 
with a radiative loss function consistent with photospheric abundances and once 
with a radiative loss function consistent with coronal abundances.  We then 
predict the expected emission measure distribution of the loop by assuming the 
loop is composed of a bundle of strands that are each in a slightly different 
stage of the temperature and plasma evolution (similar to 
\citealt{klimchuk2006,klimchuk2009,patsourakos2009}).  
We characterize the ``cool'' (6.0 $\leq$ Log T $\leq$ Log T$_{peak}$) emission 
measure distribution of each loop by assuming it can be represented by $EM \sim 
T^\alpha$ and finding $\alpha$.  

For the photospheric radiative loss function, 
we find $1.6 < \alpha < 2.0$ for all simulated loops.  For the coronal radiative 
loss function, we find $2.0 < \alpha < 2.3$ for all simulated loops. Finally, we 
sum the emission measure distributions for all loops to approximate a potential 
emission measure distribution of an active region core where many loops may be 
along the line-of-sight.  We find the summed distributions have $\alpha = 1.8$ 
for photospheric abundances and $\alpha = 2.0$ for coronal abundances.  We 
compare these results to two recent analyses of active region cores, which had 
$3.1 < \alpha < 3.4$.   We find that the long nanoflare storm scenario predicts 
$> 5$ times more 1\,MK emission than is actually observed in the 
core of these active regions.  
The long nanoflare storm also predicts more high 
temperature ($> 6$\,MK) emission than observed, but the observations are not 
well constrained at those temperatures \citep{winebarger2011b}.  
Through this 
research, we have determined that the expected index, $\alpha$, is sensitive to 
the abundances in the active region cores, with coronal abundances predicting 
larger values of $\alpha$ than photospheric abundances.  We hypothesize that if 
the abundances in the active region core were ``super-coronal'' the long 
nanoflare storm simulation could reproduce the observed emission measure 
distribution.


\section{Long Nanoflare Storm Model}

In this section, we discuss the one-dimensional hydrodynamic simulations of nine 
representative strands.  We consider three half-lengths (excluding the model 
chromosphere) of 25\,Mm, 50\,Mm and 100\,Mm.  We assume the strands are 
semi-circular, perpendicular to the solar surface, and symmetric (only the half loop
solution is solved). We 
parameterize the spatial and temporal dependence of the energy deposition as 
\begin{equation}
E_{H}(s,t) = E_0 + g(t)E_{F}\exp\left(\frac{(s-s_0)^2}{2\sigma_s^2}\right).
\label{eqn:energy}
\end{equation}
where $s_0$ designates the location of the peak of the heating function, 
$\sigma_s$ is the spatial width of the heating function, and $E_F$ determines 
the maximum amplitude of the heating function. Because symmetry is assumed,
there will be two heating locations, one along each loop leg.  
The function g(t) is chosen to be 
a simple triangular pulse,
\begin{equation}
 g(t) = \left\{ \begin{array}{ll}
             t/\delta & \mbox{$0<t\le\delta$} \\
             (2\delta-t)/\delta & \mbox{$\delta<t\le2\delta$,}
             \end{array}
       \right .
\label{eqn:gt}
\end{equation}
where $2\delta$ is the duration of the impulsive heating. The uniform background 
heating, $E_0$, consistent with $\sim 0.5$\,MK, is always applied, so the 
density and temperature will eventually return to a steady-state equilibrium.

In our simulations, the duration ($2\delta = 100$\,s), width 
($\sigma_s = 0.6$\,Mm), and heating location ($s_0 = 0.5L$ where $L$ is the 
half-length) of the impulsive  heating function is the same for each simulation. 
We vary the magnitude of the heating, $E_{F}$, to achieve a specific equilibrium 
temperature (see \citealt{winebarger2004}) of 4, 5, or 6\,MK based on the 
photospheric radiative loss function.  The equilibrium temperature in an 
impulsive heating solution is the temperature at which the density, temperature, 
and loop length all agree with steady-state conditions. It is approximately the 
temperature at which the cooling switches from being dominated by conduction to 
being dominated by radiation. This is also approximately when the density of the 
solution peaks.  The values of $E_F$ are given in Column 3 of 
Table~\ref{tab:simulation}. These same heating magnitudes are applied to the 
simulations using coronal abundances.

For each of the lengths, equilibrium temperatures, and radiative loss functions 
(18 individual simulations), we solve the hydrodynamic loop equations with the 
Naval Research Laboratory Solar Flux Tube Model (NRL\_SOLFTM). We adopt many of 
the same assumptions that were used in previous simulations with this code, and 
we refer the reader to earlier papers for additional details of this numerical 
code  (\citealt{mariska1987,mariska1989}). The radiative loss functions used in 
the simulations are based on the atomic physics calculations with ionization 
equilibrium described by \cite{mazzotta1998} and the abundances 
of \cite{grevesse1998} for photospheric abundances and \cite{feldman1992} for 
coronal abundances.

We compute average temperature and density evolution by averaging over the top 
50\% of the coronal part of the strand where $T_e > 2 \times 10^4$\,K.  As an 
example, the evolution of the 25\,Mm half-length strand with a heating 
magnitude, $E_F$, of 7.5 erg cm$^{-3}$ s$^{-1}$ is shown in Figure~\ref{fig:tn}. 
This example has an equilibrium temperature of 4\,MK. The left panel shows the 
average temperature as a function of time and the right panel shows the average 
density as a function of time for both photospheric and coronal abundances. As 
the nanoflare is turned on, the average temperature rises rapidly.  Similarly, 
the density begins to increase as chromospheric material fills the strand.  As 
the heating event ends, first the temperature and then the density begin to 
decrease. The peak density is higher and the cooling time longer for the 
photospheric solution compared to the coronal solution.

Above we have described how the temperature and density evolution of 
representative strands were calculated.  Here we discuss the method to combine 
these strands to determine properties of both individual loops and an active 
region core.  First we assume that a loop is formed of many strands, with a 
single strand in each stage of the temperature and density evolution, similar to 
\cite{klimchuk2008,patsourakos2009}.  In practical terms, this assumption 
implies a loop contains one strand for each time step of the simulation.  For 
instance, in the photospheric solution shown in Figure~\ref{fig:tn}, the strand 
evolves for 2100\,s before returning to equilibrium conditions.  The code writes 
the solution every 10\,s.  The consequent loop, then, is formed of 210 strands, 
with each strand having an average temperature and density associated with a 
specific time step of the simulation.  

The emission measure distribution of the loop is then determined by sorting the 
emission measures of the contributing strands into temperature bins. The 
emission measure for each strand is calculated from $n_e^2 ds$ where $n_e$ is 
the average electron density of the strand and $ds$ is the depth of the strand; 
this emission measure is placed in the bin containing the average temperature of 
the strand at that time step. The resulting emission measure distributions for 
each loop length, equilibrium temperature, and radiative loss assumptions are 
shown in Figure~\ref{fig:dem}. The dotted vertical lines in Figure \ref{fig:dem} 
mark the temperatures corresponding to the peak of the emission measure 
distributions; these peak temperatures are listed in Table~\ref{tab:simulation}. 
For each length and equilibrium temperature, the photospheric solutions are 
shown in blue and the coronal solutions are shown in red.  We scale the emission 
measure distributions so that the peak in the EM curve is $10^{28}$\,cm$^{-5}$.

From Log T = 6.0 to the temperature of the peak of the emission measure 
distribution, the emission measure distribution can be approximated as $EM \sim 
T^\alpha$.  We determine $\alpha$ by fitting the log of the emission measure 
distribution as a function of log of temperature with a linear regression 
routine. These fits are shown in Figure~\ref{fig:dem} and the values of $\alpha$ 
are given in Table~\ref{tab:simulation}. For photospheric abundances, $\alpha$ 
ranges from 1.6 to 2.0 and for coronal abundances $\alpha$ range from 2.0 to 
2.3.  In general, longer, hotter strands have larger $\alpha$s.  The coronal 
solutions have larger values of $\alpha$ than the photospheric solutions.

In the core of an active region, many loops are along the line-of-sight.  To 
approximate the emission measure distribution for an active region core, we 
simply sum the emission measure distributions for either the photospheric or 
coronal abundances for all loops shown in Figure~\ref{fig:dem}; these emission 
measure distributions are shown in Figure~\ref{fig:sum}. We have normalized the 
emission measure distributions to $1 \times 10^{28}$ cm$^{-5}$ for ease of 
comparison.  We also compute the $\alpha$ for the summed solutions and we find 
$\alpha = 1.8$ for photospheric abundances and $\alpha = 2.0$ for coronal 
abundances.  These $\alpha$ represent an emission measure weighted average 
$\alpha$ of the individual loops.  This implies that when an active region core 
is formed of many loops, the emission measure distribution of the core can never 
have an $\alpha$ smaller or larger than the minimum and maximum $\alpha$ for 
each constituent loop.  For comparison, we show two emission measure 
distributions from observations; we discuss these in detail in the following 
section. 

\begin{deluxetable}{ccc||cccc}
\footnotesize
\tablecaption{Simulation Results}
\tablewidth{0pt}
\tablehead{Half & Equilibrium & Volumetric & T$_{\rm Peak P}$& $\alpha_P$ & 
T$_{\rm Peak C}$  & $\alpha_C$ \\
Length & Temperature & Heating Rate, $E_F$ & (MK)  & & (MK) & \\
(Mm) & (MK) & (ergs cm$^{-3}$ s$^{-1}$) &  & & & }
\startdata
        25&         4&       7.5&       3.2&       1.6&       4.0&       2.0\\
        25&         5&      16.1&       4.0&       1.8&       5.0&       2.0\\
        25&         6&      29.5&       4.0&       2.0&       6.3&       2.2\\
        50&         4&       7.2&       3.2&       1.7&       4.0&       2.0\\
        50&         5&      14.5&       4.0&       1.9&       5.0&       2.1\\
        50&         6&      27.1&       5.0&       1.9&       5.0&       2.3\\
       100&         4&       3.3&       3.2&       1.7&       4.0&       2.0\\
       100&         5&       5.2&       4.0&       1.9&       4.0&       2.3\\
       100&         6&       8.3&       4.0&       2.0&       5.0&       2.1\\
       SUM&          &          &         3.2&       1.7&       4.0&       2.0\\

\enddata
\label{tab:simulation}
\end{deluxetable}


\section{Discussion} 

\begin{figure}[t!]
\centerline{
\resizebox{.49\textwidth}{!}{\includegraphics{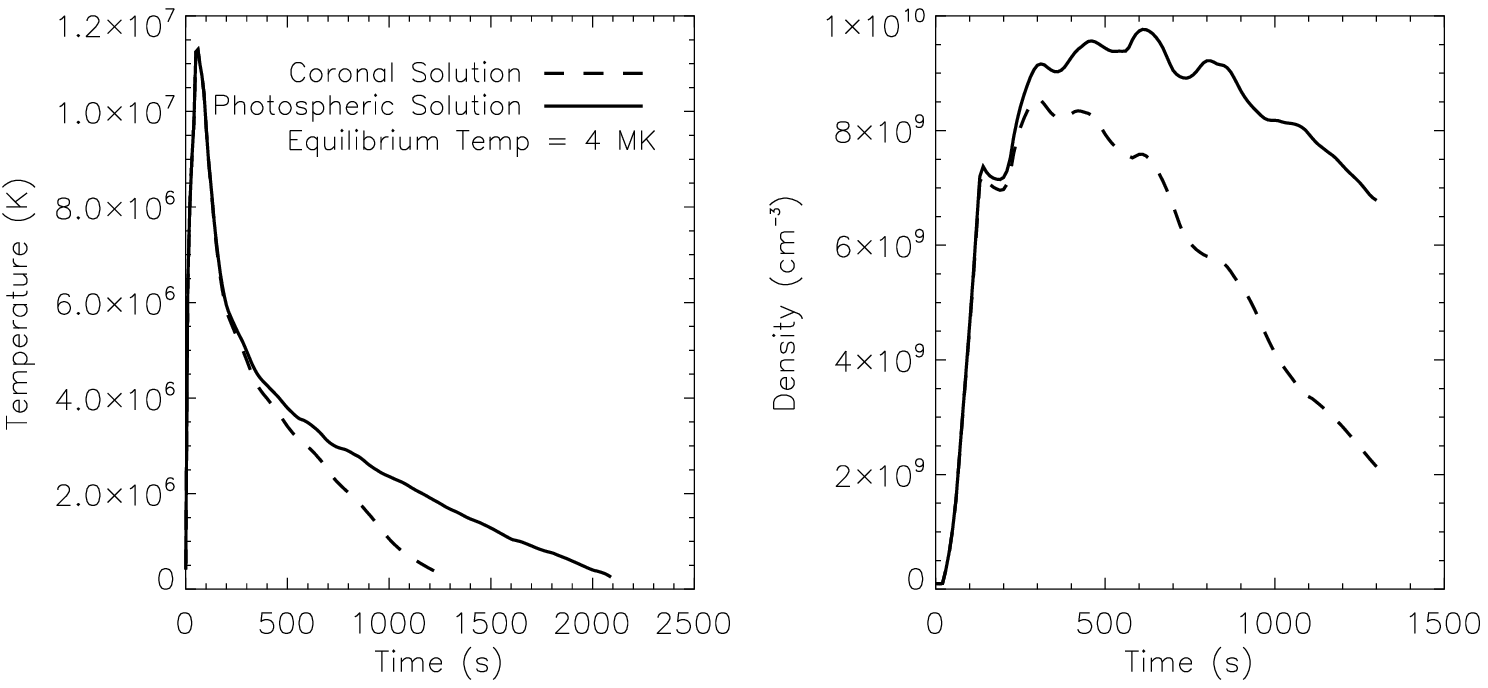}}}
\caption{The evolution of an impulsively heated, 25\,Mm half-length strand is 
shown. The average apex temperature (left) and density (right) are shown as a 
function of time for both photospheric (solid) and coronal (dashed) radiative 
loss functions.}
\label{fig:tn}
\end{figure}

In this Letter, we have investigated the emission measure distribution predicted 
by the long nanoflare storm heating scenario. We calculated an array of 
one-dimensional hydrodynamic solutions for three different lengths, three 
different equilibrium temperatures, and two assumed radiative loss functions. 
Next we computed average temperature and density evolution (Figure~\ref{fig:tn}) 
by averaging over the upper 50\% of the strand.  We assumed a loop is a bundle 
of evolving strands with each strand in a different stage of the density and 
temperature evolution and generated an emission measure distribution for each 
loop. We have calculated the power-law indices, $\alpha$, of the emission 
measure distributions for the temperature range 6.0 $<$ Log T $<$ Log T$_{peak}$ 
for the simulated loops. These measurements are listed Table 
\ref{tab:simulation}.

\begin{figure*}[t!]
\centerline{
\resizebox{\textwidth}{!}{\includegraphics{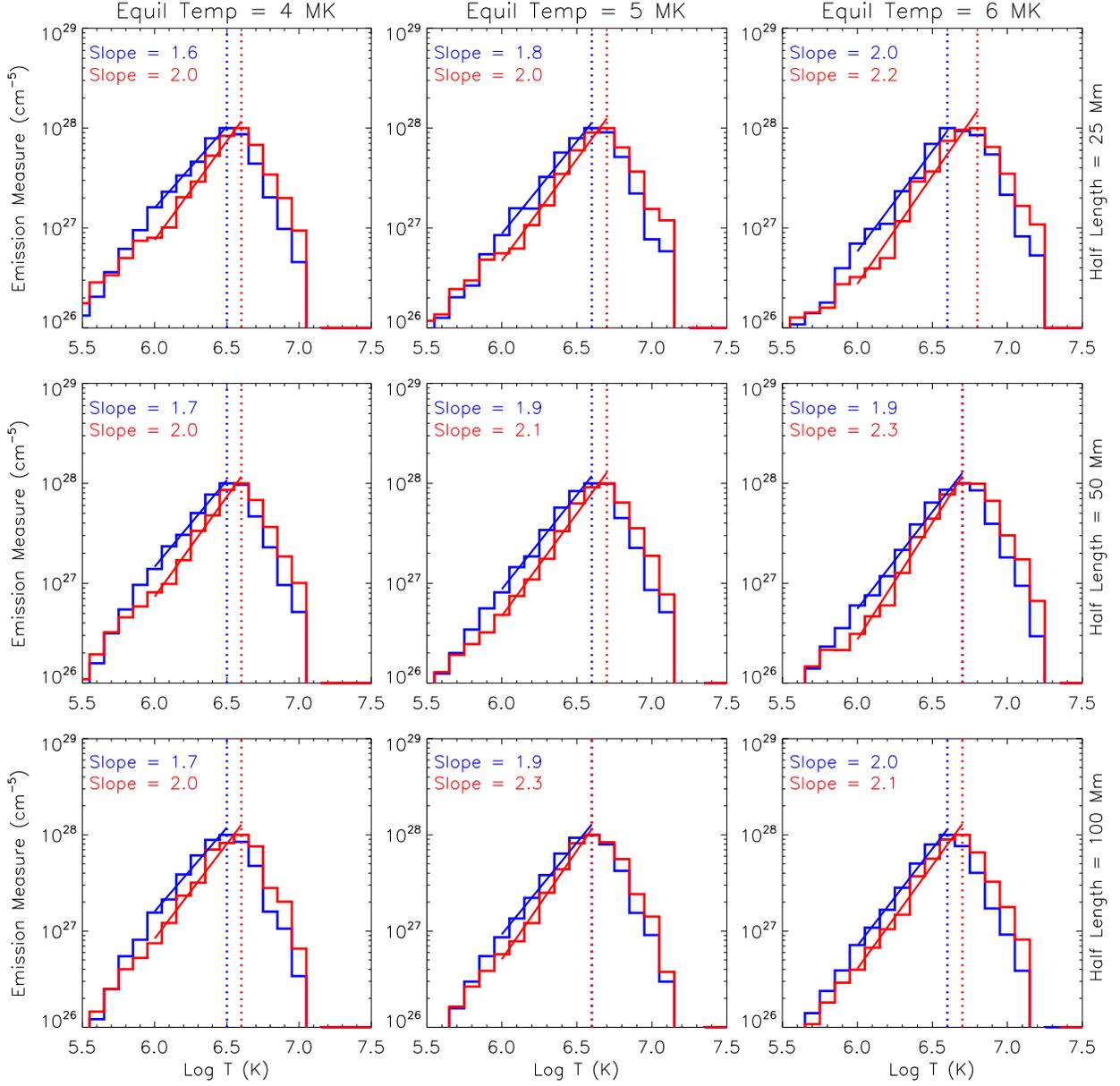}}}
\caption{The first, second and third rows show the emission measure 
distributions of 25, 50 and 100 Mm half-length loops respectively. The first, 
second, and third columns show solutions with equilibrium temperatures of 4, 5, 
and 6\,MK, respectively.  In each plot, the photospheric solution is shown as a 
blue line and the coronal solution is shown as a red line.  The fit of the 
emission measure distributions are shown as well.  The vertical dotted lines 
mark the temperatures corresponding to the peak of the emission measure 
distributions.}
\label{fig:dem}
\end{figure*}

\begin{figure}[t!]
\centerline{
\resizebox{.5\textwidth}{!}{\includegraphics{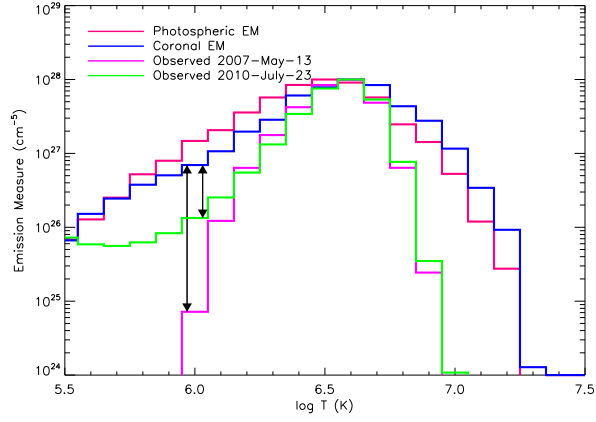}}}
\caption{The average emission measure distribution from the different loops 
shown in Figure~\ref{fig:dem}.  For comparison, two emission measure 
distributions from two recent analyses are also shown 
\cite{warren2011,winebarger2011}.  We have normalized all the emission measure 
distributions to $1 \times 10^{28}$ cm$^{-5}$.
\label{fig:sum}}
\end{figure}

To approximate an active region core, where many loops are along the 
line-of-sight, we summed the emission measure distributions that are shown in 
Figure~\ref{fig:dem}. The resulting emission measure distributions are presented 
in Figure~\ref{fig:sum}.  The power law indices of the summed emission measure 
distributions are 1.8 and 2.0 for photospheric and coronal abundances, 
respectively.  
We compare the summed emission measure distributions, shown in 
Figure~\ref{fig:sum}, to two recently published emission measure distributions 
of active region cores.  The power law indices of the observed emission measure 
distributions are $\sim 3.2$ \citep{warren2011,winebarger2011}.  When comparing 
these values with the observations, we find that the long nanoflare storm 
predicts $> 5$ times more 1\,MK emission than observed.   
We also note that the long nanoflare storm also predicts more high 
temperature ($> 6$\,MK) emission than observed (see Figure~\ref{fig:sum}), 
but we do not stress this result for several reasons.  First, the observations are not 
well constrained at high temperatures \citep{winebarger2011b}.  Also the 
high-temperature evolution of the hydrodynamic simulations, which
occurs early in the simulation when conduction dominates the energy 
equation, are greatly influenced by the location and duration of 
the heating event (see \cite{winebarger2004}).  Finally, during this early stage,
the plasma may be out of nonequilibrium ionization which would change the calculated
emission measure distribution (see \cite{bradshaw2011}).

One method of testing to see if the long nanoflare storm is viable is to examine line
ratios observed in active region cores.  Table~\ref{tab:line_rat} gives the expected
EIS intensities for a few key EIS spectral lines for the summed coronal emission measure
distribution shown in Figure~\ref{fig:sum}.  The last column of Table~\ref{tab:line_rat}
is the ratio of the expected intensities to the expected intensity in the \ion{Ca}{14} 
193.874~\AA\ line.  For instance the expected ratio between the \ion{Fe}{12} 195.119~\AA\ 
and the \ion{Ca}{14} 193.874~\AA\ lines is 13.56.  \cite{warren2011} determine the
observed ratio is closer to 4.7.  Also, the ratio between the \ion{Fe}{10} 184.536~\AA\ and \ion{Ca}{14} 193.874~\AA\
and  \ion{Fe}{15} 284.160~\AA\ and \ion{Ca}{14} 193.874~\AA\  are
predicted to be $\sim$ 3.6 and $\sim$ 86.69 respectively; the
observed ratios of these lines were $\sim 0.9$  and $\sim 33.1$
respectively in \cite{warren2011}.

\begin{deluxetable}{cccc}
\footnotesize
\tablecaption{Expected EIS intensities}
\tablewidth{0pt}
\tablehead{EIS Spectral Line & Log T$_{max}$ & $I$  &  
$I/I_{\rm Ca XIV}$\\
& & (ergs cm$^{-2}$ s$^{-1}$ sr$^{-1}$) &  }
\startdata

\ion{Fe}{10} 184.536\AA &       6.0 &     1076.9 &       3.62 \\
\ion{Fe}{11} 180.401\AA &       6.1 &     3752.2 &      12.63 \\
\ion{Fe}{11} 188.216\AA &       6.1 &     1912.2 &       6.43 \\
\ion{Fe}{12} 192.394\AA &       6.1 &     1294.2 &       4.35 \\
\ion{Fe}{12} 195.119\AA &       6.1 &     4031.3 &      13.56 \\
\ion{S}{10} 264.233\AA &        6.1 &      227.2 &       0.76 \\
\ion{Si}{10} 258.375\AA &       6.1 &     1133.9 &       3.82 \\
\ion{Fe}{13} 202.044\AA &       6.2 &     1425.0 &       4.80 \\
\ion{Fe}{13} 203.826\AA &       6.2 &     3930.6 &      13.23 \\
\ion{Fe}{14} 270.519\AA &       6.3 &     1482.4 &       4.99 \\
\ion{Fe}{14} 264.787\AA &       6.3 &     3246.6 &      10.92 \\
\ion{Fe}{15} 284.160\AA &       6.3 &    25762.0 &      86.69 \\
\ion{Fe}{16} 262.984\AA &       6.4 &     1764.9 &       5.94 \\
\ion{S}{13} 256.686\AA &        6.4 &     1684.2 &       5.67 \\
\ion{Ca}{14} 193.874\AA &       6.5 &      297.2 &       1.00 \\
\ion{Ca}{15} 200.972\AA &       6.6 &      223.6 &       0.75 \\
\ion{Ca}{16} 208.604\AA &       6.7 &      191.9 &       0.65 \\
\ion{Ca}{17} 192.858\AA &       6.7 &      424.5 &       1.43 \\

\enddata
\label{tab:line_rat}
\end{deluxetable}

We have presented emission measure distributions obtained from 
one-dimensional hydrodynamic simulations that combine plausible 
combinations of loops lengths, heating rates, and abundances 
similar to what is expected to be observed in the core of active 
regions. In early 0D nanoflare simulations, \cite{cargill1994} 
found the slope of the emission measure curve calculated from $10^6$\,K to the peak 
of the emission measure to be 3.5$ < \alpha <4.8$, though these results
are based on simple assumptions of the cooling and draining of the
plasma.

The one-dimensional hydrodynamic code used in this study can not currently deal with increasing cross sectional
area of loops and cannot be used to assess how the
expansion affects the solutions. The slope of the DEMs of loops with
expanding cross sections has been shown to get steeper with increasing
expansion factor (for example, see \citealt{vandenoord1997}) and we are working with a new code
that would be able to examine the cross sectional area in the future.


\acknowledgments
The author, FMM, is supported by an appointment to NASA's Postdoctoral
Program (NPP) which is administered by Oak Ridge Associated Universities (ORAU). The authors would like to thank the NPP host facility, Marshall Space
Flight Center, and are also grateful to the referee for providing
helpful comments to improve the manuscript.



\end{document}